\documentclass[fleqn,usenatbib]{mnras}

\usepackage{newtxtext,newtxmath}

\usepackage[T1]{fontenc}
\usepackage{ae,aecompl}
\newcommand{\caps}[1]{{\scshape{#1}}}

\usepackage{graphicx}	
\usepackage{amsmath}	
\usepackage{amssymb}	

\title[X-ray/UV Observations of SDSS J141118.31+481257.6]{\textit{Swift} X-ray and UV Observations of SDSS J141118.31+481257.6 During its First Ever Recorded Superoutburst}

\author[L.E. Rivera Sandoval]{
L.E. Rivera Sandoval$^{1}$\thanks{E-mail: liliana.rivera@ttu.edu} and
T.J. Maccarone $^{1}$
\\
$^{1}$Department of Physics and Astronomy, Box 41051, Science Building, Texas Tech University, Lubbock, TX 79409-1051, USA\\
}

\date{Accepted XXX. Received YYY; in original form ZZZ}

\pubyear{2018}

\begin{document}
\label{firstpage}
\pagerange{\pageref{firstpage}--\pageref{lastpage}}
\maketitle

\begin{abstract}
SDSS J141118.31+481257.6 is an ultracompact white dwarf binary (or AM CVn system) with an orbital period of 46 min. We analyze $\sim 23$\ ks of X-ray and UV data taken with the \textit{Neil Gehrels Swift Observatory} during its first ever recorded outbursts. The events took place 13 years after the system was discovered. We detected 3 events in our UV data, all with amplitudes of $\sim 7$ mags with respect to quiescence, the largests detected for an AM CVn system so far. 
The first 2 events correspond to a superoutburst and the third one to another detected outburst. 
The 3 episodes that we identified occurred in a period of 24 days, each one displaying very rapid brightness changes. At $\sim 120$ days since the detection of the superoutburst, the system remains 1 magnitude brighter in UV compared to the quiescence level. 
The X-ray observations suggest that the X-ray emission is not correlated with the UV.

\end{abstract}

\begin{keywords}
binaries: close--stars: individual:SDSS J141118.31+481257.6--ultraviolet: stars--X-rays: binaries--white dwarf: stars
\end{keywords}



\section{Introduction}

Ultracompact white dwarf binaries, or AM CVn systems, are binaries with orbital periods ranging from $\sim5-65$ min, with optical spectra lacking in hydrogen and dominated by helium. Given their short orbits, these systems are expected to be important sources of low frequency gravitational waves which could be detected with a mission such as \textit{LISA} \citep[e.g.][]{2018kupfer}.

Several AM CVns show outbursts, but their occurrence and recurrence rate seem to be correlated with their orbital periods ($P_{orb}$). Systems with $P_{orb}<20$ mins are in a persistent high state and do not show outbursts. Some of them do not even have an accretion disk, but accrete via direct impact instead \citep[$P_{orb}\lesssim12$ mins, see][for a review]{2010Sol}. As in the case of dwarf novae (DNe), it is believed that the outbursts in AM CVns occur due to instabilities in the accretion disk, which are triggered by changes in the opacities \citep{1974Osaki,1997tsugawa,2015Cannizzo}. However, it has been shown that other factors such as composition and changes in the mass transfer rate from the companion star ($\dot M_{tr}$) also play an important role \citep[e.g.][]{2012Kotko}.
These instabilities, and thus outbursts, are believed to occur in systems with $P_{orb}\gtrsim20$ min. The outburst recurrence time in AM CVns is very variable. Systems with $20\lesssim P_{orb}\lesssim 40$ min have recurrence times from few weeks to years \citep{2015Levitan}. Those with $40<P_{orb}<50$ experience rare outbursts (and/or rare superoutbursts) and systems with $P_{orb}\gtrsim50$ mins have not been detected in outbursts so far.

SDSS J141118.31+481257.6 is an AM CVn system \citep{2005Anderson} with an orbital period of $46\pm2$ min \citep{2007Roe} and stellar parallax of $2.361\pm0.305$\ mas \citep{2016Gaia,2018Gaia}, equivalent to a distance of $423\pm55$pc. Since its discovery, it has only been detected in quiescence, with a $g$ magnitude of 19.4 \citep{2005Anderson}. However, on 19 May 2018 Tadashi Kojima (vsnet-alert 22174\footnote{http://ooruri.kusastro.kyoto-u.ac.jp/mailarchive/vsnet-alert/22174}) reported the first ever recorded outburst of the system (with m$_V=12.4$ mags). Given the large magnitude amplitude and the presence of superhumps detected in the optical light curve by Gianluca Masi et al. 2018 (vsnet-alert 22181\footnote{http://ooruri.kusastro.kyoto-u.ac.jp/mailarchive/vsnet-alert/22181}), the event was cataloged as a superoutburst. 

AM CVns emit mostly in UV, and several of them are also X-ray sources \citep[e.g.][]{2005Ramsay,2006ramsay,2007Ramsay}. However, to date very few studies have been carried out in these bands during outbursts \citep[e.g.][]{2012ramsay}. In this paper we present the first X-ray and UV observations of SDSS J141118.31+481257.6 taken with the \textit{Neil Gehrels Swift Observatory} during the first ever recorded superoutburst of the system.



\section{Observations}

The \textit{Swift} observations considered in this paper were taken in the period from 21 May 2018 to 19 September 2018. 
They are 24 observations with a total exposure time of $\sim 23$ ks.  A typical observation had an exposure of $\sim 1$ ks (see Table \ref{data_model}). 
The X-ray and UV observations were taken simultaneously. 
For each X-ray observation there was one UV image or a series of shorter exposures. 
We report on the longest UV observation of the epoch in which the object was in focus and clearly detected.  
The UV observations were taken in the UVM2 filter ($\lambda_{cen}=2246$ \AA). For observations taken after 24 June 2018, observations in the rest of the UVOT filters (V, B, U, UVW1, UVW2) were also obtained (see Table \ref{complementary_data}). The system had B-V color of $0.2-0.3$ during these observations. There is no information in the UVM2 filter for observations performed on 21 May 2018 and 29 July 2018 due to the brightness of the object and to the total length of the exposure, respectively. However, images in the UVW1 filter were obtained in both cases ($\lambda_{cen}=2600$ \AA). 

For comparison purposes, we made use of 2 \textit{Swift} archival observations taken $\sim11$ years ago \citep{2014Evans}, when the object was in quiescence (see Tables \ref{data_model} and \ref{complementary_data}). The total length of these observations is $\sim4.6$ ks. 

To extract the UV/optical magnitudes we used the task \textsc{uvotsource} with a $5\arcsec$ circular region around the source. To estimate the UV/optical background we used a $26\arcsec$ circular region in a source-free part of the CCD. 

For the analysis of the X-ray data we first reprocessed the observations using the routine \textsc{xrtpipeline}. We determined count rates in the 0.3--10\,keV energy band using the routines \textsc{Xselect} and \textsc{Ximage}, adopting a threshold signal-to-noise ratio of 3. The rate correction factors were determined using \textsc{xrtlccorr}. Barycentric corrections were performed with the task \textsc{barycorr}. We
followed the \textit{Swift} thread\footnote{http://www.swift.ac.uk/analysis/xrt/spectra.php} for the X-ray spectral analysis.
We used \textsc{xrtmkarf} to create the ancillary response files.
For the spectral fitting we used the software package \caps{Xspec} \citep[v12.9.1,][]{1996ar}. 

Given the small number of counts in some of the individual spectra, we combined observations near the peaks of the superoutburst and outburst (m$_{UVM2}<15$) into one spectrum, and we created another one with those observations where the object was in an off-peak state ($15<$m$_{UVM2}<17$). We excluded the observations during quiescence.  
For modeling the foreground absorption, we used the model $tbnew\_gas$\footnote{http://pulsar.sternwarte.uni-erlangen.de/wilms/research/tbabs/} 
with WILM abundances \citep{2000wilm} and VERN cross-sections \citep{1996ver}. However, due to the low number of counts, we set the neutral hydrogen column, $N_H$ to the Galactic value $1.9\times10^{20}$ cm$^{-2}$. We fitted the X-ray spectra with a power-law model (\textsc{pegpwrlw}) and
with a thermal model (\textsc{vvapec}). We assumed no hydrogen and abundances as in \citet{2005Ramsay,2012ramsay}\footnote{He=3.5, C=0.04, N=12.5 and O=0.09, as appropriate for relatively low temperature CNO processed material.} for other AM CVns. The data were grouped to have 5 and 15 counts per bin (see Table \ref{fits}).
For all the fits we used the background subtracted Cash statistics \citep[W-statistics;][]{1979wa}. We used a Kolmogorov-Smirnov test to get information about the significance of the fits.

\begin{figure*}
	\includegraphics[width=14.5cm, height=15cm, trim=2cm 0 0 1cm]{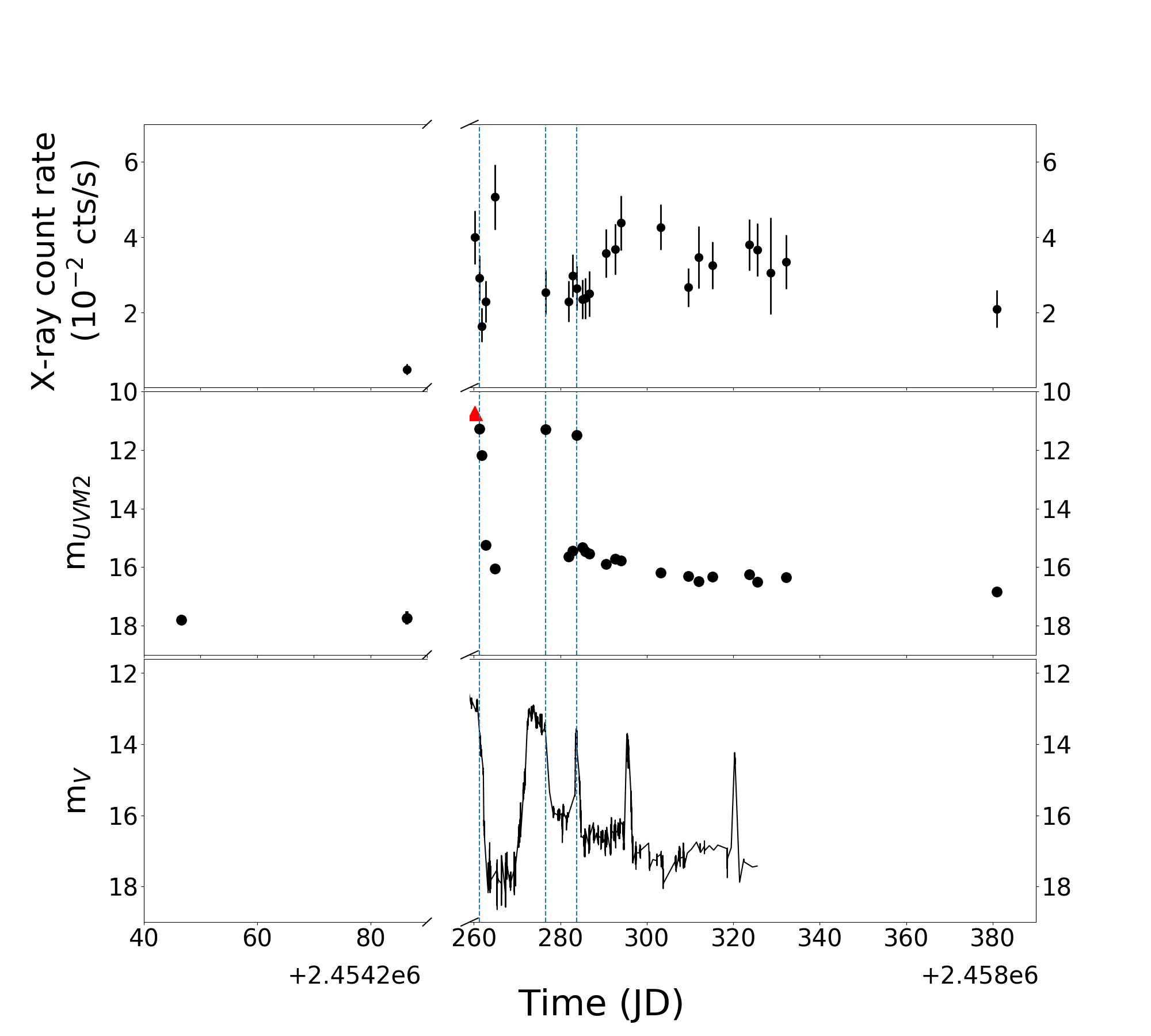}
    \caption{From top to bottom, the X-ray, UV and optical light curve of SDSS J141118.31+481257.6 during quiescence (left side) and during superoutburst/outburst (right side). The vertical lines indicate observations during the UV brightness increases for which X-ray data were also obtained. The red triangle indicates the UV lower limit for the observation on 21 May 2018. Optical data were obtained for comparison from the AAVSO data base. The quiescent optical magnitude of SDSS J141118.31+481257.6 would be below the magnitude limits in the plot ($>19$ mags, see Table \ref{complementary_data}).}
    \label{fig:lc}
\end{figure*}


\begin{figure*}
\centering
  \begin{tabular}{@{}cc@{}}
 \includegraphics[width=.5\textwidth]{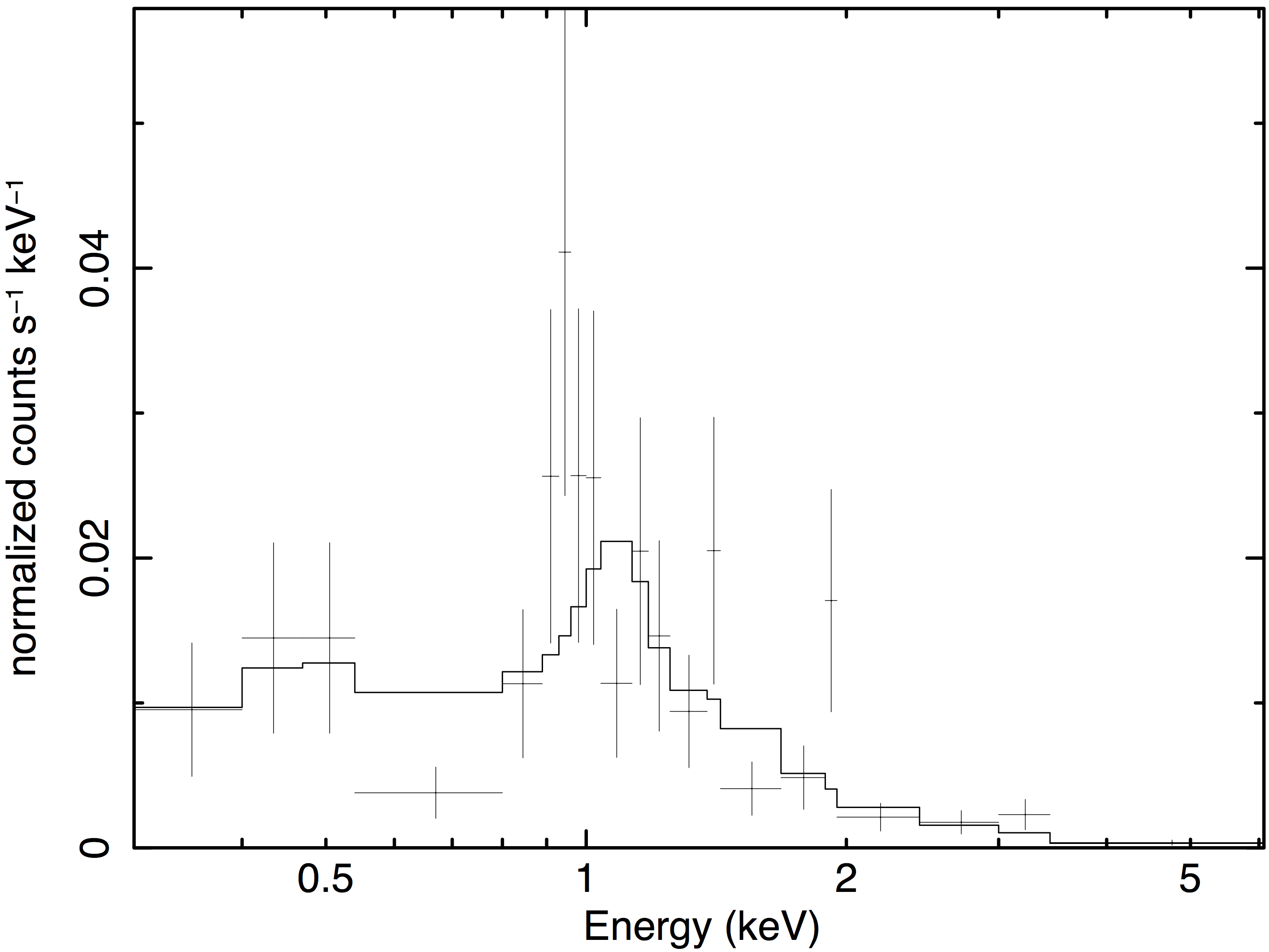}&
 \includegraphics[width=.5\textwidth]{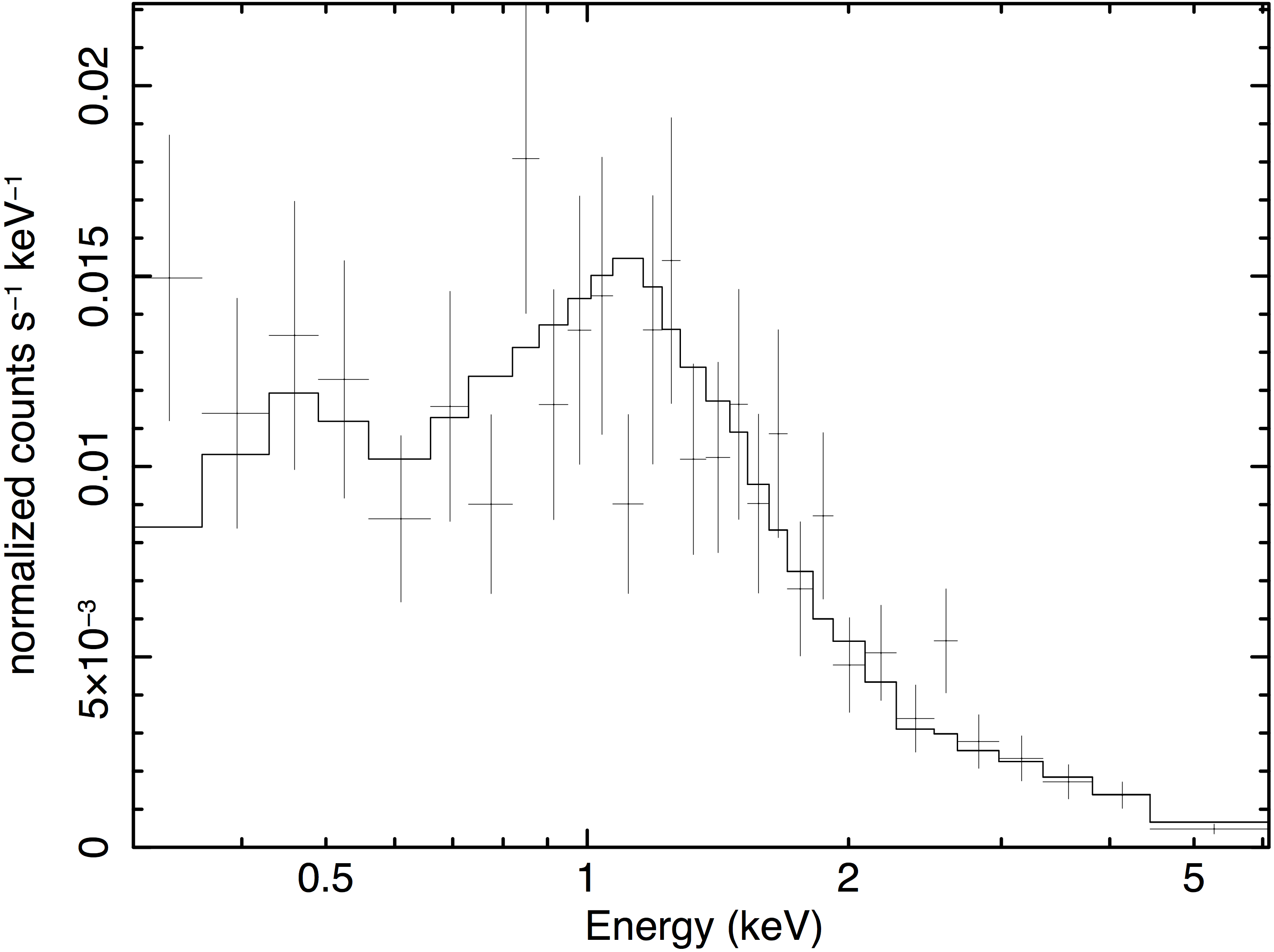}\\
  \end{tabular}
  \caption{Combined X-ray spectra of SDSS J141118.31+481257 during its active period in the 0.3--10 keV band. Left: combined spectrum for observations near the peaks (m$_{uvm2}<15$). Right: combined spectrum of observations in the off-peak state ($15<$m$_{uvm2}<17$). A thermal model is fit to each spectrum. Parameters for the fits are given in Table \ref{fits}}.
\label{fig:spectrum_out}
\end{figure*}

\section{Results}

The first \textit{Swift} observation in our data set was taken on 21 May 2018 \citep{11668}, 2 days after the discovery of the superoutburst at optical wavelengths. Because the object was brighter than the saturation limit, only an upper limit in the filter UVW1 was obtained ($<11$ mags), which means that in the bluer filter UVM2 (around which AM CVns near peak have their peak emission) the object was likely brighter. Comparing the magnitude values with archival data taken during quiescence, we determined that the superoutburst had an amplitude of at least 7 magnitudes in UV (see Fig.\ref{fig:lc}). It is the first time that such amplitude is observed in an AM CVn system 
\citep[see][for studies of other systems during superoutbursts]{2004nogami,2014kato,2015Levitan,2016isogai}.

The first peak was followed by a deep dip. Dips have been observed in other AM CVn systems during outbursts \citep{2012ramsayb} and their origin is not fully understood. However, \citet{2012Kotko} found that dips can be due to mass transfer modulations during the decay from the superoutburst maximum, which results in the propagating cooling front being caught by an outside-in heating front before the cooling reaches the inner edge of the disk. This ``catching'' would reverse the decay and the object would become bright again.

The second peak in Fig.\ref{fig:lc}, detected on 7 June 2018  \citep{11699}, would then correspond to the superouburst after the dip\footnote{Reports of superhumps with a period of 45.4 min were given by Tonny Vanmunster on 1 June 2018.}. The third peak in Fig.\ref{fig:lc} would be related to an echo outburst detected on 14 June 2018\footnote{Tonny Vanmunster also reported the presence of small humps on 13 June 2018.}.
These 2 rebrightenings detected in our UV light curve also have amplitudes of $\sim 7$ mags with respect to quiescence. 

According to the AAVSO optical light curve, 2 further echo outbursts were detected on 26 June 2018 and 20 July 2018 (see Fig.\ref{fig:lc}). Unfortunately our observations do not cover these days and thus, we will not discuss these outbursts in this paper. 

The duration of the events was variable. As can be seen in Fig.\ref{fig:lc}, the third event showed rapid increases-decreases with timescales of $\sim 24$ hrs and magnitude amplitudes of $\sim 4$ mags. The rapid bright decreases are also observed in the first event \citep{11672}, when the object decreased $\sim 3$ mags in $\sim 24$hrs. 

In Figure \ref{fig:lc}, we also show the results of our X-ray observations in the energy band 0.3--10 keV. Unfortunately the low number of counts obtained in each observation leads to large errors. However, while we see obvious brightness increases in the UV light curve, the X-ray count rates do not suggest the same behavior. In Figure \ref{fig:spectrum_out} we present the combined spectra of the individual X-ray observations near the peak and in the off-peak state. We fitted the absorbed thermal model (\textsc{vvapec}) to determine the temperature of the plasma, as well as the absorbed power-law model to get information about the power-law index ($\Gamma$) in both states. Results of the fits are given in Table \ref{fits}. No significant differences were observed in $\Gamma$ in both states. However, the amount of data to perform a good fit is not large, especially for observations near the peak. An slightly increase in the X-ray flux is observed for the observations in the off-peak state, which suggests that in that state the disk became optically thin and thus, more X-ray photons were detected. 

Unfortunately, the number of X-ray counts during quiescence is very low and no good constrains can be derived with any model. Therefore, we do not discuss the X-ray spectra from these observations.

\begin{table*}
\centering
\caption{Parameter values of the X-ray spectral fits of SDSS J141118.31+481257.6 near the peak of the superoutbursts and in the off-peak state. Observations for each state were divided according to their UV magnitude as indicated in the first column of this table. Unabsorbed X-ray luminosities ($L_{X_{pl}}$) were obtained with the power-law model and assuming a distance of $423\pm55$pc. Errors in $L_{X_{pl}}$ are dominated by the errors in distance. Goodness of the fits were calculated by using a KS test. All errors are given at 90\% confidence level.}
\label{fits}
\begin{tabular}{l c c c c c c c}
State & cts/bin & $\Gamma$ & $F_{X_{pl}}$ &  $L_{X_{pl}}$  &  Goodness & kT   & Goodness \\   
&               &          & ($\times 10^{-12}$ erg cm$^{-2}$ s$^{-1}$) & ($\times 10^{31}$ erg s$^{-1}$) & (PL) &(keV) & (Thermal) \\
\hline
\hline
Near-peak (m$_{uvm2}<15$)         & 5 & $1.8\pm 0.2$ & $0.95^{+0.17}_{-0.22}$ & $2.0^{+0.9}_{-1.0}$ & 99.8\% & $2.3^{+1.6}_{-0.8}$ & 74.9\%\\
Off-peak ($15<$m$_{uvm2}<17$) & 15& $1.6\pm 0.1$ & $1.29\pm 0.14$         & $2.8 \pm 1.0$  & 82.7\% & $5.9^{+2.8}_{-1.5}$ & 39.6\% \\
\end{tabular}
\end{table*}

\section{discussion}

SDSS J141118.31+481257.6 has been identified in superoutburst for the first time since its discovery 13 years ago. Its UV lightcurve shows the largest magnitude amplitude of an AM CVn system in superoutbursts detected so far. 

The masses of the components are unknown and our data do not allow us to obtain an estimation of them (e.g. by estimating the mass ratio through period excess, though there are relatively large uncertainties on $P_{orb}$). However, from the relation between $P_{orb}$ and the companion mass given by \citet{2015Cannizzo}, we can infer that the companion star in SDSS J141118.31+481257.6 should have a mass $\sim 0.015$ M$_\odot$. Additionally, from evolutionary models of AM CVns \citep[Fig.3\footnote{Models for that figure were provided by Chris Deloye.} of][]{2012Kotko}, 
the mass transfer rate of the companion in a 46 min system is $\sim1.6\times 10^{14}$ g s$^{-1}$, which means that using Eq. 14 of \citet[][]{2012Kotko}, the mass of the primary star has to be $\gtrsim0.45$ M$_\odot$, otherwise the disk would be cold and stable, and no outbursts would have been detected. 

The UV light curve and the long quiescent period of the system would favor a low mass for the primary star. From Fig. \ref{fig:lc} we see that after the peaks, the object remains in a less bright state, which is $\sim 1-1.5$ mags brighter than the quiescent level. This can be understood if we take into account that when the mass of the primary star is small, the inner disk radius is larger. Thus, in order to reach the critical surface density necessary to trigger the instabilities, the object has to accumulate first a larger amount of mass. As a consequence, the quiescent period becomes long, which agrees with the long quiescent period in SDSS J141118.31+481257.6 (at least 13 years). Once the object is in (super)outburst, the time to reach quiescence will also be longer given the large amount of mass accumulated in the disk that has to be accreted. This will continue until the system reaches the critical surface density to make the disk stable again. All these changes are perhaps related to drastic changes in $\dot M_{tr}$
(the enhanced mass-transfer model initially proposed by \citet[][]{1983Vogt} and later discussed/applied by \citet{2012Kotko} to AM CVns). 
However, the tidal-thermal instability model, which among other characteristics helps to explain the presence of growing superhumps from the precursor to the main superoutburst in short period DNe \citep[see e.g.][]{2013Osaki}, could also be involved in the presence of superoutbursts in AM CVns. 
If changes in $\dot M_{tr}$ are involved in the case of  SDSS J141118.31+481257.6, these have to occur 
in very short periods of time ($\sim 24$ h), as dramatic bright increases and decays were observed.

SDSS J141118.31+481257.6 has been above the quiescent level for at least $\sim 60$ days after the last outburst (considering the AAVSO event on 20 July 2018 as the last one). Additional to the mechanism discussed in the last paragraph, it is possible that a large amount of heat was deposited in the WD during the superoutburst, which would require a larger amount of time to be released. Therefore, the object would look brighter for a longer period after accretion has stopped. This effect has been observed on DNe. For example, \citet{1996gansicke} and  \citet{1996Sion,2001Sion} found that after a superoutburst the WD is hotter and cools slower than after a normal outburst. 

The X-ray count rates of SDSS J141118.31+481257.6 do not seem to be correlated with the brightness increases in UV or optical. This could be the result of the boundary layer becoming optically thick to X-rays during the superoutbursts. In fact, the power-law fit suggests that the X-ray flux slightly increased during the off-peak observations, which would correspond to the boundary layer becoming optically thin. This behavior has previously been observed in other AM CVns in outbursts \citep[e.g. KL Dra][]{2010ramsay,2012ramsay} and also on outbursting DNe. 
For instance, the DN SS Cyg displays a similar behavior in optical, UV and X-rays \citep{2003Wheatley}. This can be explained when the accretion rate exceeds a critical value making the boundary layer optically thick during outburst, ``switching the X-rays off" and emitting mostly in extreme UV \citep{2003Schreiber}. During outburst the optical and UV emission of SS Cyg would be dominated by the accretion disk.

The X-ray spectra of SDSS J141118.31+481257.6 during the peak and off-peak states seem to be similarly hard.
There is also weak evidence for a cooler plasma near the peak of the superoutburst but a similar behavior has also been suggested for KL Dra \citep[][]{2012ramsay}. We noted that in the near-peak spectrum there is a bin at $\sim0.65$ keV with a remarkable low number of counts. This could be due to an ionized O absorption line. However, the errors in the spectrum are large and thus no good constrains can be obtained. 

The superoutburst amplitude, frequency and decay time-length of SDSS J141118.31+481257.6 will help to constrain new AM CVn models, as up to now the statistics of systems in superoutburst with $P_{orb}$ longer than 37 min is not large\footnote{It is believed that systems with $P_{orb} \gtrsim 37$ min have very long superoutburst recurrence times, analogous to WZ Sge stars \citep{2015Levitan}.}. Unfortunately to obtain more data for this and similar systems (long period AM CVns with one or very few superoutbursts already observed) is not easy given their long recurrence times.  Furthermore, available relations for the duration, magnitude amplitude and recurrence time in AM CVns (useful to plan observations), like those derived by \citet{2015Levitan} cannot be fully applied to long period systems, as they have been derived for relatively short period AM CVns ($P_{orb} \lesssim 37$min) and for which several superoutbursts have been detected. However, if we use these relations \citep[Sec. 4.2 of][]{2015Levitan} to get a rough estimate of the superoutburst duration in SDSS J141118.31+481257.6, we obtain that it would be $\sim 100$ days. This value is much larger than the observed duration in the UV or optical light curves (see fig.\ref{fig:lc}), which would be $\sim20$ days\footnote{Though, the exact duration is uncertain due to the lack of observations days before the first report of the superoutburst.} including the dip. However, we can split the event into 2 superoutbursts. Therefore, if the second rebrightening is indeed due to enhanced mass overflow from the secondary star due to irradiation from the first superoutbursts, the duration would be $\sim10$ days. In both cases the observed duration of the event would be well below the predicted value, suggesting a less steep $P_{orb}$ vs. outburst duration relation than that of \citet{2015Levitan}.

\section*{Acknowledgements}
We acknowledge the comments of the anonymous referee, which improved this manuscript.
We thank the \textit{Swift} team for approving and scheduling our multiple target of opportunity requests, which provided us with the data used in this paper. Additional thanks to Tadashi Kojima, Patrick Schmeer and Tonny Vanmunster for their reports regarding this object. We also thank Yuri Cavecchi for comments on the manuscript. We acknowledge with thanks the variable star observations from the AAVSO International Database contributed by observers worldwide and used in this research.
This work has made use of data from the European Space Agency (ESA) mission
{\it Gaia} (\url{https://www.cosmos.esa.int/gaia}), processed by the {\it Gaia}
Data Processing and Analysis Consortium (DPAC,
\url{https://www.cosmos.esa.int/web/gaia/dpac/consortium}). Funding for the DPAC
has been provided by national institutions, in particular the institutions
participating in the {\it Gaia} Multilateral Agreement.



\bibliographystyle{mnras}
\bibliography{biblio} 

\begin{thebibliography}{}
\makeatletter
\relax
\def\mn@urlcharsother{\let\do\@makeother \do\$\do\&\do\#\do\^\do\_\do\%\do\~}
\def\mn@doi{\begingroup\mn@urlcharsother \@ifnextchar [ {\mn@doi@}
  {\mn@doi@[]}}
\def\mn@doi@[#1]#2{\def\@tempa{#1}\ifx\@tempa\@empty \href
  {http://dx.doi.org/#2} {doi:#2}\else \href {http://dx.doi.org/#2} {#1}\fi
  \endgroup}
\def\mn@eprint#1#2{\mn@eprint@#1:#2::\@nil}
\def\mn@eprint@arXiv#1{\href {http://arxiv.org/abs/#1} {{\tt arXiv:#1}}}
\def\mn@eprint@dblp#1{\href {http://dblp.uni-trier.de/rec/bibtex/#1.xml}
  {dblp:#1}}
\def\mn@eprint@#1:#2:#3:#4\@nil{\def\@tempa {#1}\def\@tempb {#2}\def\@tempc
  {#3}\ifx \@tempc \@empty \let \@tempc \@tempb \let \@tempb \@tempa \fi \ifx
  \@tempb \@empty \def\@tempb {arXiv}\fi \@ifundefined
  {mn@eprint@\@tempb}{\@tempb:\@tempc}{\expandafter \expandafter \csname
  mn@eprint@\@tempb\endcsname \expandafter{\@tempc}}}

\bibitem[\protect\citeauthoryear{{Anderson} et~al.,}{{Anderson}
  et~al.}{2005}]{2005Anderson}
{Anderson} S.~F.,  et~al., 2005, \mn@doi [\aj] {10.1086/491587}, \href
  {http://adsabs.harvard.edu/abs/2005AJ....130.2230A} {130, 2230}

\bibitem[\protect\citeauthoryear{{Arnaud}}{{Arnaud}}{1996}]{1996ar}
{Arnaud} K.~A.,  1996, in {Jacoby} G.~H.,  {Barnes} J.,  eds,  Astronomical
  Society of the Pacific Conference Series Vol. 101, Astronomical Data Analysis
  Software and Systems V. p.~17

\bibitem[\protect\citeauthoryear{{Cannizzo} \& {Nelemans}}{{Cannizzo} \&
  {Nelemans}}{2015}]{2015Cannizzo}
{Cannizzo} J.~K.,  {Nelemans} G.,  2015, \mn@doi [\apj]
  {10.1088/0004-637X/803/1/19}, \href
  {http://adsabs.harvard.edu/abs/2015ApJ...803...19C} {803, 19}

\bibitem[\protect\citeauthoryear{{Evans} et~al.,}{{Evans}
  et~al.}{2014}]{2014Evans}
{Evans} P.~A.,  et~al., 2014, \mn@doi [\apjs] {10.1088/0067-0049/210/1/8},
  \href {http://adsabs.harvard.edu/abs/2014ApJS..210....8E} {210, 8}

\bibitem[\protect\citeauthoryear{{Gaia Collaboration} et~al.,}{{Gaia
  Collaboration} et~al.}{2016}]{2016Gaia}
{Gaia Collaboration} et~al., 2016, \mn@doi [\aap]
  {10.1051/0004-6361/201629272}, \href
  {http://adsabs.harvard.edu/abs/2016A%26A...595A...1G} {595, A1}

\bibitem[\protect\citeauthoryear{{Gaia Collaboration} et~al.,}{{Gaia
  Collaboration} et~al.}{2018}]{2018Gaia}
{Gaia Collaboration} et~al., 2018, \mn@doi [\aap]
  {10.1051/0004-6361/201833051}, \href
  {http://adsabs.harvard.edu/abs/2018A%26A...616A...1G} {616, A1}

\bibitem[\protect\citeauthoryear{{G{\"a}nsicke} \& {Beuermann}}{{G{\"a}nsicke}
  \& {Beuermann}}{1996}]{1996gansicke}
{G{\"a}nsicke} B.~T.,  {Beuermann} K.,  1996, \aap, \href
  {http://adsabs.harvard.edu/abs/1996A%26A...309L..47G} {309, L47}

\bibitem[\protect\citeauthoryear{{Isogai}, {Kato}, {Ohshima}, {Kasai},
  {Oksanen}, {Masumoto}  \& {Fukushima}}{{Isogai} et~al.}{2016}]{2016isogai}
{Isogai} K.,  {Kato} T.,  {Ohshima} T.,  {Kasai} K.,  {Oksanen} A.,  {Masumoto}
  K.,   {Fukushima} D.,  2016, \mn@doi [\pasj] {10.1093/pasj/psw063}, \href
  {http://adsabs.harvard.edu/abs/2016PASJ...68...64I} {68, 64}

\bibitem[\protect\citeauthoryear{{Kato} et~al.,}{{Kato}
  et~al.}{2014}]{2014kato}
{Kato} T.,  et~al., 2014, \mn@doi [\pasj] {10.1093/pasj/psu077}, \href
  {http://adsabs.harvard.edu/abs/2014PASJ...66L...7K} {66, L7}

\bibitem[\protect\citeauthoryear{{Kotko}, {Lasota}, {Dubus}  \&
  {Hameury}}{{Kotko} et~al.}{2012}]{2012Kotko}
{Kotko} I.,  {Lasota} J.-P.,  {Dubus} G.,   {Hameury} J.-M.,  2012, \mn@doi
  [\aap] {10.1051/0004-6361/201219156}, \href
  {http://adsabs.harvard.edu/abs/2012A%26A...544A..13K} {544, A13}

\bibitem[\protect\citeauthoryear{{Kupfer} et~al.,}{{Kupfer}
  et~al.}{2018}]{2018kupfer}
{Kupfer} T.,  et~al., 2018, \mn@doi [\mnras] {10.1093/mnras/sty1545}, \href
  {http://adsabs.harvard.edu/abs/2018MNRAS.480..302K} {480, 302}

\bibitem[\protect\citeauthoryear{{Levitan}, {Groot}, {Prince}, {Kulkarni},
  {Laher}, {Ofek}, {Sesar}  \& {Surace}}{{Levitan} et~al.}{2015}]{2015Levitan}
{Levitan} D.,  {Groot} P.~J.,  {Prince} T.~A.,  {Kulkarni} S.~R.,  {Laher} R.,
  {Ofek} E.~O.,  {Sesar} B.,   {Surace} J.,  2015, \mn@doi [\mnras]
  {10.1093/mnras/stu2105}, \href
  {http://adsabs.harvard.edu/abs/2015MNRAS.446..391L} {446, 391}

\bibitem[\protect\citeauthoryear{{Nogami}, {Monard}, {Retter}, {Liu}, {Uemura},
  {Ishioka}, {Imada}  \& {Kato}}{{Nogami} et~al.}{2004}]{2004nogami}
{Nogami} D.,  {Monard} B.,  {Retter} A.,  {Liu} A.,  {Uemura} M.,  {Ishioka}
  R.,  {Imada} A.,   {Kato} T.,  2004, \mn@doi [\pasj] {10.1093/pasj/56.6.L39},
  \href {http://adsabs.harvard.edu/abs/2004PASJ...56L..39N} {56, L39}

\bibitem[\protect\citeauthoryear{{Osaki}}{{Osaki}}{1974}]{1974Osaki}
{Osaki} Y.,  1974, \pasj, \href
  {http://adsabs.harvard.edu/abs/1974PASJ...26..429O} {26, 429}

\bibitem[\protect\citeauthoryear{{Osaki} \& {Kato}}{{Osaki} \&
  {Kato}}{2013}]{2013Osaki}
{Osaki} Y.,  {Kato} T.,  2013, \mn@doi [\pasj] {10.1093/pasj/65.3.50}, \href
  {http://adsabs.harvard.edu/abs/2013PASJ...65...50O} {65, 50}

\bibitem[\protect\citeauthoryear{{Ramsay}, {Hakala}, {Marsh}, {Nelemans},
  {Steeghs}  \& {Cropper}}{{Ramsay} et~al.}{2005}]{2005Ramsay}
{Ramsay} G.,  {Hakala} P.,  {Marsh} T.,  {Nelemans} G.,  {Steeghs} D.,
  {Cropper} M.,  2005, \mn@doi [\aap] {10.1051/0004-6361:20052950}, \href
  {http://adsabs.harvard.edu/abs/2005A%26A...440..675R} {440, 675}

\bibitem[\protect\citeauthoryear{{Ramsay}, {Groot}, {Marsh}, {Nelemans},
  {Steeghs}  \& {Hakala}}{{Ramsay} et~al.}{2006}]{2006ramsay}
{Ramsay} G.,  {Groot} P.~J.,  {Marsh} T.,  {Nelemans} G.,  {Steeghs} D.,
  {Hakala} P.,  2006, \mn@doi [\aap] {10.1051/0004-6361:20065491}, \href
  {http://adsabs.harvard.edu/abs/2006A%26A...457..623R} {457, 623}

\bibitem[\protect\citeauthoryear{{Ramsay} et~al.,}{{Ramsay}
  et~al.}{2007}]{2007Ramsay}
{Ramsay} G.,  et~al., 2007, in {Napiwotzki} R.,  {Burleigh} M.~R.,  eds,
  Astronomical Society of the Pacific Conference Series Vol. 372, 15th European
  Workshop on White Dwarfs. p.~425 (\mn@eprint {} {astro-ph/0610357})

\bibitem[\protect\citeauthoryear{{Ramsay} et~al.,}{{Ramsay}
  et~al.}{2010}]{2010ramsay}
{Ramsay} G.,  et~al., 2010, \mn@doi [\mnras]
  {10.1111/j.1365-2966.2010.17019.x}, \href
  {http://adsabs.harvard.edu/abs/2010MNRAS.407.1819R} {407, 1819}

\bibitem[\protect\citeauthoryear{{Ramsay}, {Barclay}, {Steeghs}, {Wheatley},
  {Hakala}, {Kotko}  \& {Rosen}}{{Ramsay} et~al.}{2012a}]{2012ramsayb}
{Ramsay} G.,  {Barclay} T.,  {Steeghs} D.,  {Wheatley} P.~J.,  {Hakala} P.,
  {Kotko} I.,   {Rosen} S.,  2012a, \mn@doi [\mnras]
  {10.1111/j.1365-2966.2011.19924.x}, \href
  {http://adsabs.harvard.edu/abs/2012MNRAS.419.2836R} {419, 2836}

\bibitem[\protect\citeauthoryear{{Ramsay}, {Wheatley}, {Rosen}, {Barclay}  \&
  {Steeghs}}{{Ramsay} et~al.}{2012b}]{2012ramsay}
{Ramsay} G.,  {Wheatley} P.~J.,  {Rosen} S.,  {Barclay} T.,   {Steeghs} D.,
  2012b, \mn@doi [\mnras] {10.1111/j.1365-2966.2012.21660.x}, \href
  {http://adsabs.harvard.edu/abs/2012MNRAS.425.1486R} {425, 1486}

\bibitem[\protect\citeauthoryear{{Rivera Sandoval} \& {Maccarone}}{{Rivera
  Sandoval} \& {Maccarone}}{2018a}]{11668}
{Rivera Sandoval} L.~E.,  {Maccarone} T.,  2018a, The Astronomer's Telegram,
  \href {http://adsabs.harvard.edu/abs/2018ATel11668....1R} {11668}

\bibitem[\protect\citeauthoryear{{Rivera Sandoval} \& {Maccarone}}{{Rivera
  Sandoval} \& {Maccarone}}{2018b}]{11672}
{Rivera Sandoval} L.~E.,  {Maccarone} T.,  2018b, The Astronomer's Telegram,
  \href {http://adsabs.harvard.edu/abs/2018ATel11672....1R} {11672}

\bibitem[\protect\citeauthoryear{{Rivera Sandoval} \& {Maccarone}}{{Rivera
  Sandoval} \& {Maccarone}}{2018c}]{11699}
{Rivera Sandoval} L.~E.,  {Maccarone} T.,  2018c, The Astronomer's Telegram,
  \href {http://adsabs.harvard.edu/abs/2018ATel11699....1R} {11699}

\bibitem[\protect\citeauthoryear{{Roelofs}, {Groot}, {Steeghs}, {Marsh}  \&
  {Nelemans}}{{Roelofs} et~al.}{2007}]{2007Roe}
{Roelofs} G.~H.~A.,  {Groot} P.~J.,  {Steeghs} D.,  {Marsh} T.~R.,   {Nelemans}
  G.,  2007, \mn@doi [\mnras] {10.1111/j.1365-2966.2007.12397.x}, \href
  {http://adsabs.harvard.edu/abs/2007MNRAS.382.1643R} {382, 1643}

\bibitem[\protect\citeauthoryear{{Schreiber}, {Hameury}  \&
  {Lasota}}{{Schreiber} et~al.}{2003}]{2003Schreiber}
{Schreiber} M.~R.,  {Hameury} J.-M.,   {Lasota} J.-P.,  2003, \mn@doi [\aap]
  {10.1051/0004-6361:20031221}, \href
  {http://adsabs.harvard.edu/abs/2003A%26A...410..239S} {410, 239}

\bibitem[\protect\citeauthoryear{{Sion}, {Cheng}, {Huang}, {Hubeny}  \&
  {Szkody}}{{Sion} et~al.}{1996}]{1996Sion}
{Sion} E.~M.,  {Cheng} F.-H.,  {Huang} M.,  {Hubeny} I.,   {Szkody} P.,  1996,
  \mn@doi [\apjl] {10.1086/310318}, \href
  {http://adsabs.harvard.edu/abs/1996ApJ...471L..41S} {471, L41}

\bibitem[\protect\citeauthoryear{{Sion}, {Cheng}, {Szkody}, {G{\"a}nsicke},
  {Sparks}  \& {Hubeny}}{{Sion} et~al.}{2001}]{2001Sion}
{Sion} E.~M.,  {Cheng} F.-H.,  {Szkody} P.,  {G{\"a}nsicke} B.,  {Sparks}
  W.~M.,   {Hubeny} I.,  2001, \mn@doi [\apjl] {10.1086/324558}, \href
  {http://adsabs.harvard.edu/abs/2001ApJ...561L.127S} {561, L127}

\bibitem[\protect\citeauthoryear{{Solheim}}{{Solheim}}{2010}]{2010Sol}
{Solheim} J.-E.,  2010, \mn@doi [\pasp] {10.1086/656680}, \href
  {http://adsabs.harvard.edu/abs/2010PASP..122.1133S} {122, 1133}

\bibitem[\protect\citeauthoryear{{Tsugawa} \& {Osaki}}{{Tsugawa} \&
  {Osaki}}{1997}]{1997tsugawa}
{Tsugawa} M.,  {Osaki} Y.,  1997, \mn@doi [\pasj] {10.1093/pasj/49.1.75}, \href
  {http://adsabs.harvard.edu/abs/1997PASJ...49...75T} {49, 75}

\bibitem[\protect\citeauthoryear{{Verner}, {Ferland}, {Korista}  \&
  {Yakovlev}}{{Verner} et~al.}{1996}]{1996ver}
{Verner} D.~A.,  {Ferland} G.~J.,  {Korista} K.~T.,   {Yakovlev} D.~G.,  1996,
  \mn@doi [\apj] {10.1086/177435}, \href
  {http://adsabs.harvard.edu/abs/1996ApJ...465..487V} {465, 487}

\bibitem[\protect\citeauthoryear{{Vogt}}{{Vogt}}{1983}]{1983Vogt}
{Vogt} N.,  1983, \aap, \href
  {http://adsabs.harvard.edu/abs/1983A%26A...118...95V} {118, 95}

\bibitem[\protect\citeauthoryear{{Wachter}, {Leach}  \& {Kellogg}}{{Wachter}
  et~al.}{1979}]{1979wa}
{Wachter} K.,  {Leach} R.,   {Kellogg} E.,  1979, \mn@doi [\apj]
  {10.1086/157084}, \href {http://adsabs.harvard.edu/abs/1979ApJ...230..274W}
  {230, 274}

\bibitem[\protect\citeauthoryear{{Wheatley}, {Mauche}  \& {Mattei}}{{Wheatley}
  et~al.}{2003}]{2003Wheatley}
{Wheatley} P.~J.,  {Mauche} C.~W.,   {Mattei} J.~A.,  2003, \mn@doi [\mnras]
  {10.1046/j.1365-8711.2003.06936.x}, \href
  {http://adsabs.harvard.edu/abs/2003MNRAS.345...49W} {345, 49}

\bibitem[\protect\citeauthoryear{{Wilms}, {Allen}  \& {McCray}}{{Wilms}
  et~al.}{2000}]{2000wilm}
{Wilms} J.,  {Allen} A.,   {McCray} R.,  2000, \mn@doi [\apj] {10.1086/317016},
  \href {http://adsabs.harvard.edu/abs/2000ApJ...542..914W} {542, 914}

\makeatother
\end{thebibliography}



\appendix

\section{\textit{Swift} X-ray and UV/optical data }

In the following tables we provide the UV and X-ray data behind Figure \ref{fig:lc} for SDSS J141118.31+481257 during its superoutburst period. Additional UV and optical data were obtained for several observations during the off-peak state. 

\begin{table*}
\centering
\caption{\textit{Swift} data for SDSS J141118.31+481257. The first 2 entries correspond to archival observations when the object was in quiescence. Apparent magnitudes are given in Vega system, the provided errors in magnitudes include photometric and systematic errors. Exposure times are given for the X-ray and UV (UVM2) observations, respectively. X-ray count rates are given in the 0.3--10 keV band. Dates have been barycentered. Magnitudes are not derredened.}
\label{data_model}
\begin{tabular}{c c c c  }
Date & UVM2 & Exposure  &  X-ray count rate              \\               
(JD) &      & X-ray/UVM2 (s) &  ($\times 10^{-2}$ cts/s) \\
\hline
\hline
2454246.59483796 & 17.8(2)  & 1490 (116) & -\\
2454286.48771991 & 17.7(2)  & 3094 (68)  & $0.5 \pm 0.1$\\
\hline
2458260.25681713 & $<11.0$  & 956 (-)    & $3.9 \pm 0.7$ \\
2458261.23608796 & 11.28(4) & 976 (768)  & $2.9 \pm 0.6$ \\
2458261.84071759 & 12.18(4) & 976 (794)  & $1.6 \pm 0.5$ \\
2458262.70206019 & 15.25(5) & 974 (300)  & $2.2 \pm 0.5$ \\
2458264.88983796 & 16.06(5) & 874 (715)  & $5.1 \pm 0.8$ \\
2458276.57986111 & 11.29(4) & 927 (287)  & $2.5 \pm 0.6$ \\
2458281.89756944 & 15.64(4) & 1019 (811) & $2.3 \pm 0.5$ \\
2458282.88151620 & 15.44(6) & 1124 (302) & $3.0 \pm 0.6$ \\
2458283.75324074 & 11.49(4) & 946 (795)  & $2.6 \pm 0.6$ \\
2458285.13799769 & 15.32(4) & 1049 (854) & $2.3 \pm 0.5$ \\
2458285.73589120 & 15.46(4) & 986 (806) & $2.4 \pm 0.5$ \\
2458286.73445602 & 15.55(4) & 931 (772)  & $2.5 \pm 0.6$ \\
2458290.58288194 & 15.90(5) & 1049 (831) & $3.6 \pm 0.6$ \\
2458292.64158565 & 15.72(5) & 1011 (815) & $3.7 \pm 0.7$ \\
2458293.97621528 & 15.78(5) & 1004 (845) & $4.4 \pm 0.7$ \\
2458303.20143519 & 16.19(9) & 1411 (123) & $4.2 \pm 0.6$ \\
2458309.65508102 & 16.31(9) & 1209 (126) & $2.7 \pm 0.5$ \\
2458312.03957176 & 16.5(1)  & 607 (105)  & $3.4 \pm 0.8$ \\
2458315.16525463 & 16.32(9) & 956 (151)  & $3.2 \pm 0.6$ \\
2458323.66343750 & 16.26(9) & 999 (123) & $3.8 \pm 0.7$ \\
2458325.52539352 & 16.50(8) & 936 (208)  & $3.7 \pm 0.7$ \\
2458328.57391204 &  -       & 217 (-)    & $3.0 \pm 1.4$ \\
2458332.22652778 & 16.34(8) & 747 (178)  & $3.3 \pm 0.7$ \\
2458380.97969907 & 16.84(9) & 971 (242)  & $2.1 \pm 0.5$\\
\end{tabular}
\end{table*}

\begin{table*}
\centering
\caption{\textit{Swift} UV and optical data of SDSS J141118.31+481257.6 during quiescence(first 2 entries) and for several off-peak observations. Apparent magnitudes are given in Vega system. Reported errors include photometric and systematic errors. The given dates correspond to those of the X-ray observations. Magnitudes are not derredened.}
\label{complementary_data}
\begin{tabular}{c c c c c c  }
Date & UVW2 & UVW1  & U & B & V  \\               
(JD) &      &       &   &   &    \\
\hline
\hline
2454246.59483796 & 17.9(2) & 18.0(2) & 18.3(1) & $>19$ & $>19$ \\
2454286.48771991 & 18.2(2) & 17.9(2) & 17.9(2) & $>18$ & $>18.4$\\
\hline
2458303.20143519 & 16.17(7) & 16.18(7) & 16.32(6) & 17.30(6) & 17.1(2)\\
2458309.65508102 & 16.28(8) & 16.17(7) &  16.41(6) & 17.31(7) & 17.0(1)\\
2458312.03957176 &  16.44(9) & 16.27(9) & 16.34(5) & 17.30(7) & 17.0(1)\\
2458315.16525463 & 16.27(8) & 16.15(7) & 16.35(6) & 17.34(6) & 17.11(8)\\
2458323.66343750 & 16.42(8) & 16.25(9) & 16.8(1) & 17.8(2) & 17.8(3) \\
2458325.52539352 & 16.50(7) & 16.43(8) & 17.0(1) & 17.9(1) & 17.9(2) \\
2458328.57391204 &    --    & 16.63(9) & 17.0(1) &   --    &   --    \\
2458332.22652778 & 16.37(7) & 16.34(9) & 16.76(9) & 17.7(1) & 17.4(2) \\
2458380.97969907 & 16.77(6) & 16.84(9) & 17.21(9) & 18.0(1) & 17.8(2)\\
\end{tabular}
\end{table*}


\bsp	
\label{lastpage}
\end{document}